\documentclass[pra,showpacs,11pt,superscriptaddress,floatfix,showkeys]{revtex4-1}
\usepackage{graphicx}
\usepackage{amsmath}
\usepackage{epstopdf}
\usepackage[usenames,dvipsnames,svgnames,table]{xcolor}
\usepackage{upgreek}

\begin{document}

\newcommand{\degree}{^\circ}
\newcommand{\unit}[1]{\,\mathrm{#1}}
\newcommand{\Figureref}[2][]{Figure~\ref{#2}#1}
\newcommand{\Eqnref}[1]{Equation~(\ref{#1})}
\newcommand{\bra}[1]{\left<\mathrm{#1}\right|}
\newcommand{\ket}[1]{\left|\mathrm{#1}\right>}

\newcommand{\affilOU}{Homer L. Dodge Department of Physics and Astronomy, The University of Oklahoma, 440 W. Brooks St. Norman, OK 73019, USA}
\newcommand{\affilStutt}{5. Physikalisches Institut, Universit\"{a}t Stuttgart, Pfaffenwaldring 57 D-70550 Stuttgart, Germany}

\title{Rydberg-atom based radio-frequency electrometry using frequency modulation spectroscopy in room temperature vapor cells}
\author{Santosh Kumar}
\affiliation{\affilOU}
\author{Haoquan Fan}
\affiliation{\affilOU}
\author{Harald K\"{u}bler}
\affiliation{\affilStutt}
\author{Akbar J. Jahangiri}
\affiliation{\affilOU}
\author{James P. Shaffer}
\affiliation{\affilOU}
\email{Corresponding author: shaffer@nhn.ou.edu}
%\affiliation{Homer L. Dodge Department of Physics and Astronomy, The University of Oklahoma, 440 W. Brooks St. Norman, OK 73019, USA}
%\affilStutt

\date{\today}
\begin{abstract}
Rydberg atom-based electrometry enables traceable electric field measurements with high sensitivity over a large frequency range, from gigahertz to terahertz. Such measurements are particularly useful for the calibration of radio frequency and terahertz devices, as well as other applications like near field imaging of electric fields. We utilize frequency modulated spectroscopy with active control of residual amplitude modulation to improve the signal to noise ratio of the optical readout of Rydberg atom-based radio frequency electrometry. Matched filtering of the signal is also implemented. Although we have reached similarly, high sensitivity with other read-out methods, frequency modulated spectroscopy is advantageous because it is well-suited for building a compact, portable sensor. In the current experiment, $\sim$3 $\mu$V cm$^{\mathrm{-1}}$ Hz$^{\mathrm{-1/2}}$ sensitivity is achieved and is found to be photon shot noise limited. %and require a new procedure to improve the sensitivity using frequency modulation spectroscopy.
\end{abstract}

\maketitle

\section{Introduction}
Atoms and molecules have been used as precision standards for measurements of fundamental physical quantities such as time, length, mass, and the gravitational constant \cite{Weiss93,Cronin09,Rosi14,Ludlow15,Degen16}. The measurement of electric and magnetic fields has also been improved by using atoms as sensors \cite{Booth94,Crowley04,Paulusse05,Savukov05,Obrecht07,Mohapatra07,Mohapatra08,Bohi10,Balabas10,Dolde11,Sedlacek12,Bohi12,Acosta13,1Holloway14,Fan15,Sedlacek16,Naber16}. Radio frequency (RF) electric field (E-field) measurements are important because the RF spectrum is the foundation of modern communications \cite{Pozar01}, remote sensing \cite{Davis78} and many other applications \cite{Thostenson99}, including some in medical science \cite{Rosen02}. Recently, atom-based sensors have been used for RF E-field measurements using the concept of electromagnetically induced transparency (EIT) with Rydberg atoms in room temperature vapor cells \cite{Sedlacek12,Fan16}. Rydberg atoms in vapor cells can also be used for quantum information science \cite{Mohapatra08,Kubler10,Pfau09,Pfau11}. RF E-fields can be used to coherently manipulate Rydberg atoms for these purposes \cite{Barato14} which significantly increases the scientific interest in studying the interaction between RF E-fields and atoms in vapor cells \cite{Kubler10,Urvoy15,Fanapp}.

Rydberg atom-based RF E-field measurements are superior to conventional antenna-based techniques and can be used, in principle, over the range of GHz-THz. It has been shown that Rydberg atom-based methods have surpassed the sensitivity limit, $\mathrm{\sim 1\,  mV cm^{-1}Hz^{-1/2}}$, and accuracies, $4-20\%$ depending on frequency, of current dipole antenna-based traceable RF E-field measurements \cite{Tishchenko03}.  Rydberg atom-based electrometry can also reach frequencies above $\sim 150\,$GHz for which there is no current traceable standard \cite{Gordon14}. We have demonstrated in our prior work that it is possible to achieve a sensitivity of $\sim$ 5 $\mu$Vcm$^{-1}$Hz$^{-1}$ and accuracies of $\sim 1\%$ using Rydberg atom-based sensors at GHz frequencies \cite{Sedlacek12,Kumar16}. Imaging with sub-wavelength resolution \cite{Fan14, 2Holloway14} and vector RF E-field measurements \cite{Sedlacek13} can be used to extend the utility of the technique. In this paper, we study frequency modulated (FM) spectroscopy as a low noise and, in principle, compact way to read-out the probe laser signal of the EIT system that carries the information about the RF E-field. This study is important for pushing the sensitivity of the technique to the atomic projection noise limit by eliminating technical noise in the probe laser read-out and elucidating the role of probe laser shot noise. The projection noise limit of the Rydberg atom-based E-field measurement can be on the order of $\sim$pV$\,$cm$^{-1}$ \cite{Fan15}.

\begin{figure}[htbp]
  \begin{center}
\includegraphics[width=3.0 in]{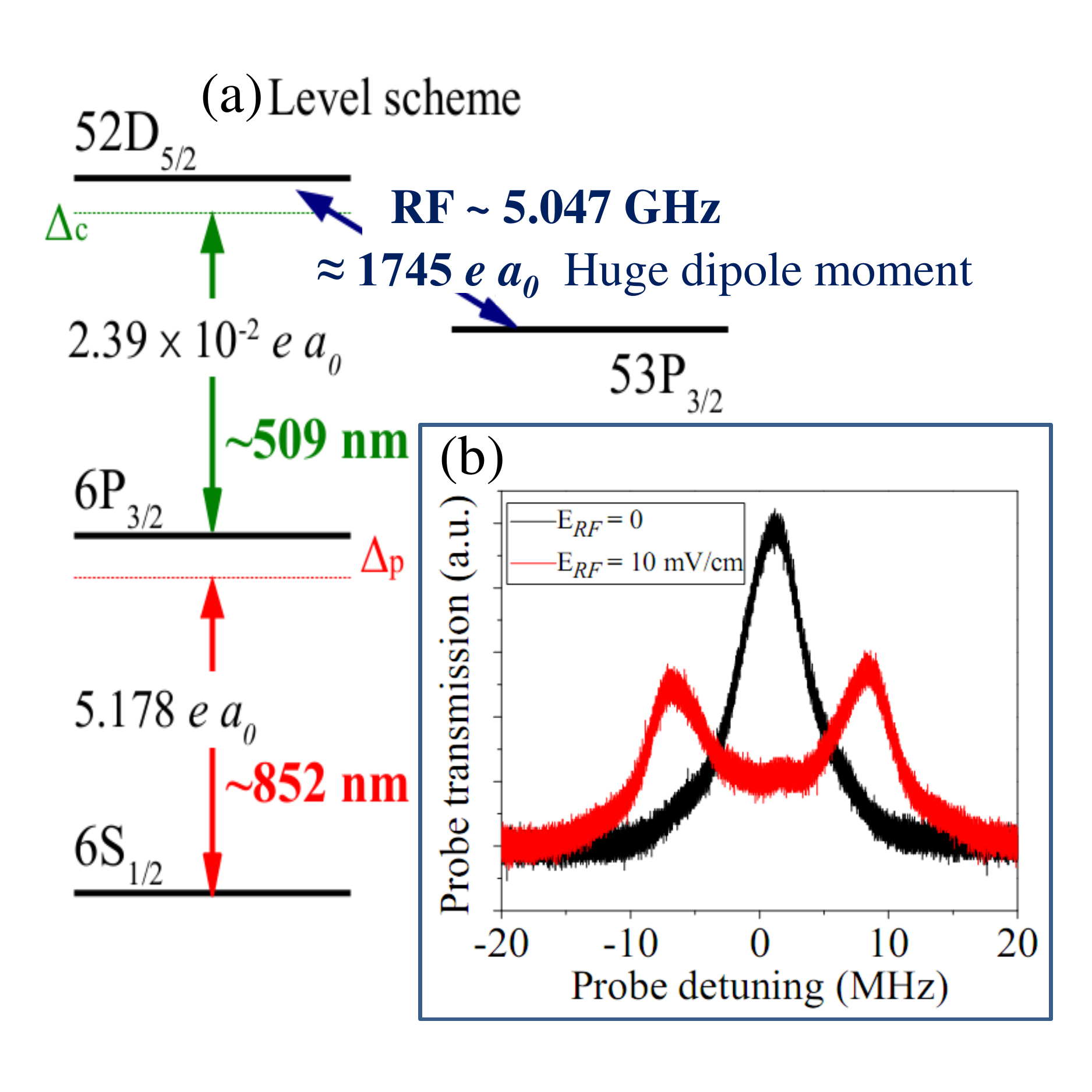}   % requires the graphicx packageQCMC
  \end{center}
   \caption{ (a) The atomic energy level scheme for the measurements. The probe laser has a wavelength of $\sim 852\,$nm while the coupling laser wavelength is $\sim 509\,$nm. These lasers are nearly resonant with both transitions. (b) Shows the probe transmission signal vs probe laser detuning for different RF E-field strengths. These measurements show the AT regime. The measurements were carried out with amplitude modulation of the coupling laser beam similar to \cite{Sedlacek12}.}
   \label{fig0}
\end{figure}

At the heart of Rydberg atom-based RF electrometry is the large transition dipole moment, $\mu_{RF}$, between two adjacent Rydberg states, e.g. 100 to 10000 times the alkali D$_2$ transition \cite{Gallagher}, which allows one to detect absolute RF E-fields with high sensitivity. When the RF E-field is at the level of mV$\,$cm$^{-1}$, Autler-Townes (AT) splitting is generated in the EIT spectrum by the RF E-field coupling between the Rydberg states, Fig.~\ref{fig0} \cite{Fan15, Gordon14}. The AT splitting, $\Delta \nu$, is proportional to the transition dipole moment between the Rydberg states and the RF E-field amplitude, i.e. $\Delta \nu \propto \mu_{RF} E_{RF}/h$, where $h$ is Planck's constant. The principle quantum number, $n$, of the Rydberg states can be changed to cover a broad frequency spectrum from GHz to THz, with varying sensitivity depending primarily on the transition dipole moments and collision rates. When the AT splitting cannot be observed due to the spectral resolution, the amplitude of the EIT probe transmission can be used to determine the E-field \cite{Sedlacek12}. The latter measurement is more sensitive to noise because the probe transmission on resonance and/or the spectral lineshape must be accurately determined. To increase the sensitivity, two goals present themselves. First, it is important to increase the signal-to-noise ratio (SNR) of the measurement. Second, it is advantageous to increase the spectral resolution to expand the AT regime. This paper addresses the first goal using a method, FM spectroscopy, that can be integrated into a portable, compact package.

%Measurement of magnetic and electric fields \cite{koschorreck2010sub,jensen2010quantum,balabas2010polarized,savukov2005tunable}, can also be improved with atom-based methods. Atom-based standards have the advantages of being linked to precision measurements of fundamental constants and atomic, or molecular, properties \cite{weiss1993precision}.The atoms are partially excited to Rydberg states to determine the amplitude of a RF E-field. and demonstrated the detection of weak RF E-field $\sim$ 8 $\mu$V/cm As a result, Rydberg atom-based RF E-field sensing is a promising candidate for a new standard for RF  E-fields. Our results agreed with the Doppler averaged density matrix calculations.

The sensitivity of the RF E-field measurement is limited by three main factors from reaching the projection noise limit of the atomic sensor. These factors are: the technical noise of the probe laser, the residual Doppler effect due to the wavelength mismatch of the EIT probe and coupling lasers, and the photon shot noise of the probe laser on the photodetector. Recently, we have utilized a homodyne detection technique with a Mach-Zehnder interferometer (MZI) to suppress the technical noise of the probe laser \cite{Kumar16}. We achieved a new sensitivity limit for atom-based RF E-field sensing of $\sim \mathrm{5\, \mu V cm^{-1} Hz^{-1/2} }$ which was determined to be photon shot noise limited.  The MZI is experimentally more complicated than FM spectroscopy because it requires an additional frequency stabilized laser to control the phase of the interferometer. FM spectroscopy is an alternative, less complex method, that can reduce the technical noise of the probe laser so that photon shot noise limited performance can be realized \cite{Bjorklund80,Bjorklund81}.
% and open up a new regime to utilize it to make traceable RF E-field measurements.

%However, a straightforward analysis of the photon shot noise on the photodector implies that improvements in the coupling laser noise and interferometer stability will only be incremental.

%We present progress on overcoming the residual Doppler effect by using a new multi-photon scheme and reaching the shot noise detection limit using frequency modulated spectroscopy. Our experiments also show promise for studying quantum optical effects such as superradiance in vapor cells using Rydberg atoms.

FM spectroscopy has been used for precision measurements such as gravitational wave detection interferometry \cite{Kokeyama14,Vine09}, cavity-based laser frequency stabilization \cite{Black01,Yu16}, and high-speed detection of weak absorption signals \cite{Riris96,Ye03}. This method has also been used to suppress light shifts in optical pumping systems \cite{McGuyer09}. In order to achieve high sensitivity with FM spectroscopy residual amplitude modulation (RAM) must be addressed. RAM is caused by etalon effects, variations of the birefringence of the electro-optic modulator (EOM) used for frequency modulation, fluctuations in the RF drive power, spatial variation of the field inside the EOM crystal, etc that lead to unequal magnitude and/or unwanted phase shifts of the modulation sidebands. Active control and cancellation of RAM improves the readout SNR and is required to optimize FM spectroscopy \cite{Wong85,Zhang14}. Recently, an alternative method has also been proposed to passively control RAM. A wedged electro-optic crystal was used to reduce RAM caused by the input polarization misalignment and the etalon effects \cite{Li16}. This method is useful for the reduction of baseline drift of the Pound--Drever--Hall (PDH) error signal for stable squeezed light generation. A wedged electro-optic crystal is not required with active control of the RAM.

In this work, we show that we can achieve a sensitivity of $\sim \mathrm{3\, \mu V cm^{-1} Hz^{-1/2} }$ for Rydberg atom-based E-field measurement with FM spectroscopy. We actively stabilize the RAM in our setup. Our sensitivity using FM spectroscopy is also photon shot noise limited and virtually the same as that obtained with the MZI. These results support our work using the MZI and show that photon shot noise is a challenge for Rydberg atom-based E-field measurements since it is difficult to increase the probe laser power without increasing collision rates and introducing unwanted amounts of power broadening. %This method is a passive control of the RAM using the wedged shape of the crystal.

%($\sim$5 $\mu$V cm$^{\mathrm{-1}}$ Hz$^{\mathrm{-1/2}})$
%In this paper, we show that  the frequency modulation spectroscopy with active control of residual amplitude modulation improves the read out signal to noise 20 times. The method has the potential for high sensitivity ($\sim$3 $\mu$V cm$^{\mathrm{-1}}$ Hz$^{\mathrm{-1/2}})$ and can be self-calibrated.

\section{Experimental Method}
Fig.~\ref{fig0}(a) shows the four-level atomic energy level scheme we used for the experiments. The probe laser is tuned near the cesium (Cs) D$_2$ transition, $6S_{1/2} (F=4)\rightarrow 6P_{3/2} (F'=5)$, while the coupling laser excites atoms to a Rydberg state, $6P_{3/2} (F'=5)\rightarrow 52 D_{5/2}$. The three-level ladder system with the couplings shown in Fig.~\ref{fig0} leads to the cancellation of absorption of a resonant probe laser in the presence of a resonant coupling laser. The enhanced transmission of the probe laser is known as EIT. A RF E-field at a frequency of $5.047\,$GHz, resonant with two adjacent Rydberg levels, $52D_{5/2} \leftrightarrow 53P_{3/2}$, can cause the probe transmission window to split via the AT effect with proper selection of laser polarization \cite{Sedlacek12,Sedlacek13}. The RF E-field is determined by how it modifies the probe laser transmission. Fig.~\ref{fig0}(b) shows the typical probe transmission signal as a function of probe laser detuning in the AT regime. The traces shown in Fig.~\ref{fig0} were obtained using amplitude modulation of the coupling laser as used for our previous works \cite{Sedlacek12,Fan16,Kumar16}.

\begin{figure}[htbp]
  \begin{center}
\includegraphics[width=4in]{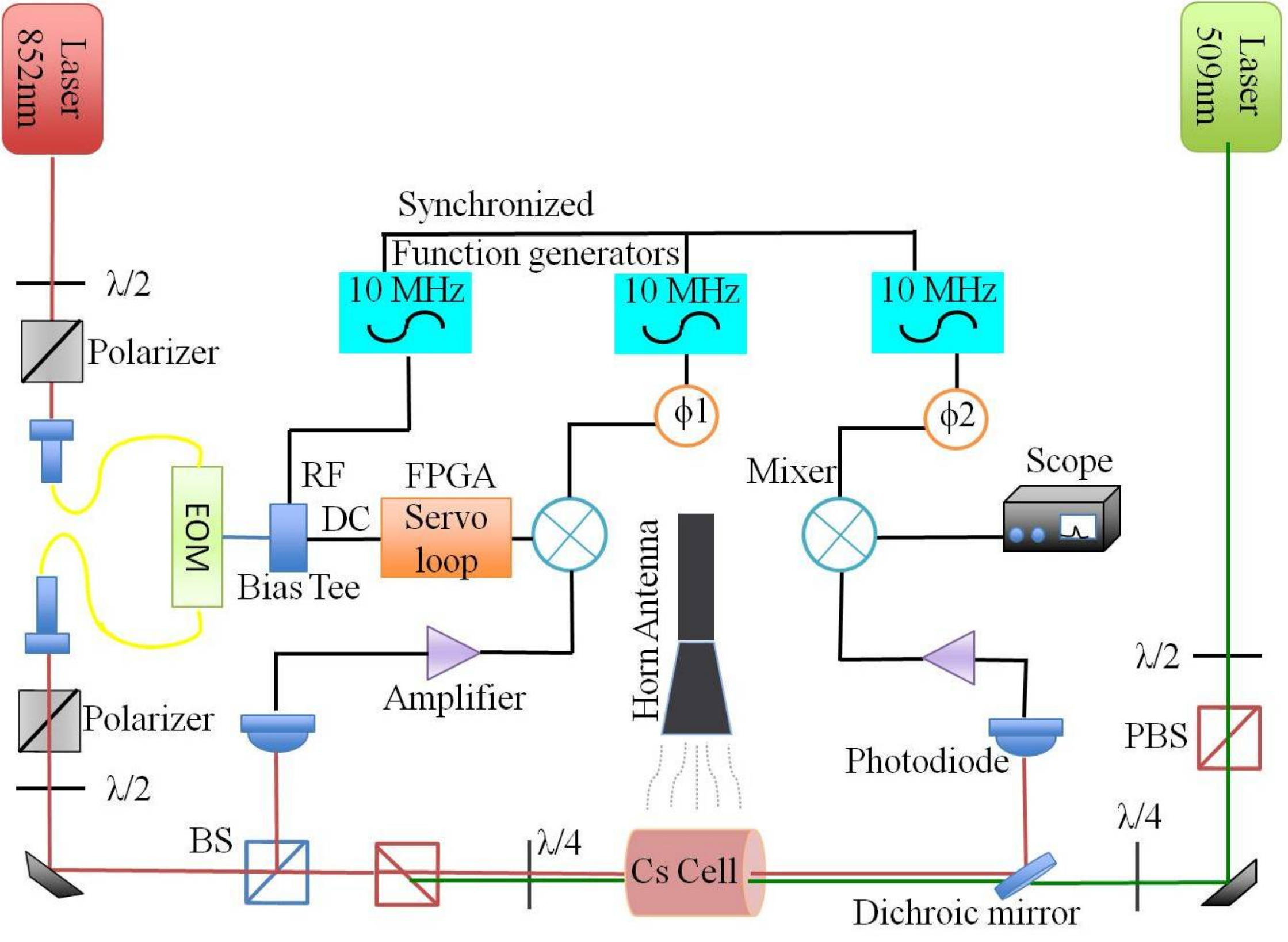}   % requires the graphicx packageQCMC
  \end{center}
   \caption{Shows a schematic of the experimental setup for the FM spectroscopy experiments, where acronyms are SAS, saturation absorption spectroscopy; PDH, Pound--Drever--Hall locking; EOM, electro-optic phase modulator; AOM, acousto-optic modulator; GTP, Glan-Thompson prism; BS, beam splitter; PBS, polarizing beam splitter; FPGA, field programming gate array; $\lambda/2 (\lambda/4)$ half (quarter) -waveplate; PD, photodetector; BPF, band pass filter; LPF, low pass filter; $\phi_1, \phi_2$, phase shifters; DBM, double balanced mixer; IOS, optical isolator; DM, dichroic mirror.}
   \label{fig1}
\end{figure}

The experimental setup for the implementation of FM spectroscopy is shown in Fig.~\ref{fig1}. A $3\,$cm long, $1\,$cm$^2$ cross-sectional area rectangular vapor cell filled with Cs atoms at room temperature is used to perform the RF E-field measurements. The lasers are offset locked to an ultra-stable Fabry--P\'erot cavity using the PDH technique \cite{Black01}. The estimated linewidth of the lasers is $\sim 50\,$kHz based on the cavity locking error signal. The two laser beams interact with the Cs atoms in a counter-propagating geometry. The intensity fluctuations of both lasers are stabilized using a feedback loop to acousto-optic modulators (AOM) for the sensitivity measurements. RF absorbing material is placed around the setup to suppress reflections of the RF E-field.

To modulate the probe laser, a linear polarized probe laser beam is incident on a fiber-coupled waveguide-based EOM with small angle relative to the crystal axis. To produce high purity linear polarized light, two Glan-Thompson prisms (GTP) are placed at the input and output of the EOM to act as a polarizer and analyzer, respectively. The modulation frequency that we used for the experiments is $10\,$MHz which is larger than the spectral width of the EIT probe transmission window.

%This AT splitting is proportional to the amplitude of the RF E-field \cite{Sedlacek12}. The AT frequency splitting, $\mathrm{\Delta}$, is
%\begin{equation}
%\mathrm{\Delta=\frac{\lambda_{c}}{\lambda_{p}}\frac{\mu_{RF} E_{RF}}{h}}, \label{eqn:rf}
%\end{equation}
%where $\mathrm{\lambda_{p}}$ and $\mathrm{\lambda_{c}}$ are the wavelengths of the probe and coupling lasers, respectively. $\mathrm{\mu_{RF}}$ is the transition dipole moment, $\mathrm{E}$ is the amplitude of the RF E-field, and $\mathrm{h}$ is Planck's constant. In Eq. (\ref{eqn:rf}), we also take into account the residual Doppler effect due to the wavelength mismatch ($\mathrm{\lambda_{c}/\lambda_{p}}$) of the counter-propagating lasers \cite{mohapatra2007coherent,sedlacek2012microwave}.

%The lockin is used to detect the EIT signal using AM modulation of coupling beam.
RAM causes the FM sidebands of the probe laser to vary in relative magnitude and phase. RAM can be actively compensated by rotating the principal axes of the EOM crystal by the application of an electric field in order to align the linear polarized probe laser with one of the principal axes of the crystal. By aligning the probe laser polarization with one of the principal axes of the crystal, the probe light propagates as if in a homogeneous material.  A waveguide-based EOM is advantageous because the applied voltages needed to produce the necessary optical phase shift are 2 orders of magnitude smaller and the input fiber acts as a spatial mode filter, which reduces the RAM contributions arising from the spatial inhomogeneity of the optical field \cite{Zhang14}. We temperature stabilized the waveguide-based EOM to reduce drift due to the temperature fluctuations.

\begin{figure*}[htbp]
  \begin{center}
 \includegraphics[width=2.6in]{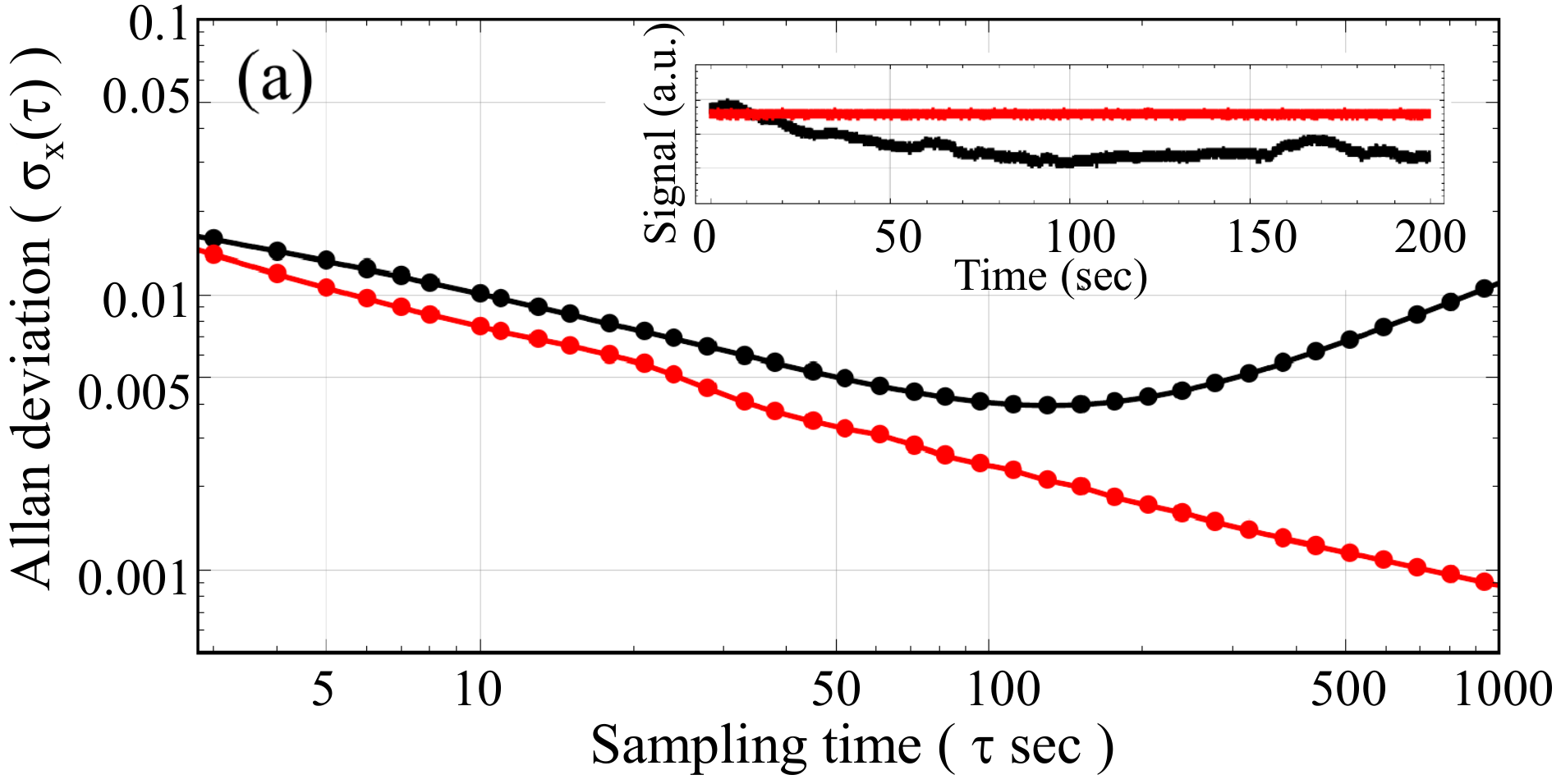}
  \includegraphics[width=2.5in]{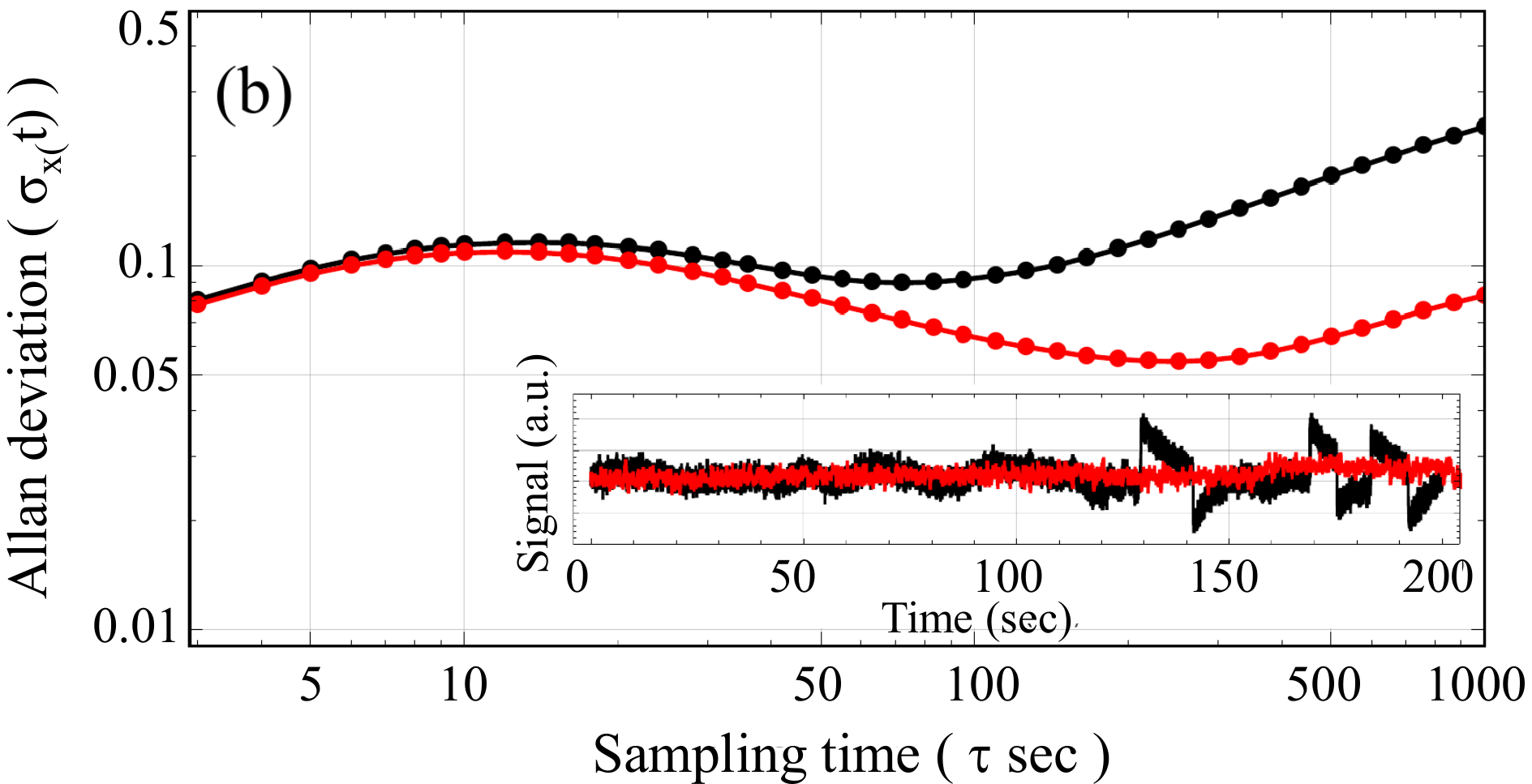}% requires the graphicx packageQCMC
  \end{center}
   \caption{ %(a) Shows the change in RAM signal vs applied voltage. (a) Shows the RAM signal vs time with lock on (red) and off (black), and (b) shows the FFT of the RAM signal with lock on (red) and off (black).
   (a) Shows an example of the Allan deviation of the RAM signal vs. sampling time with the RAM lock on (red) and off (black). Inset of (a) shows the corresponding RAM signal vs. time with the RAM lock on (red) and off (black). (b) Shows an example of the Allan deviation of the probe laser transmission signal on the EIT resonance vs. sampling time with the RAM lock on (red) and off (black). Inset of (b) shows the corresponding FM signal at two-photon resonance vs. time with the RAM lock on (red) and off (black). The units are arbitrary for these plots because they were calculated directly from the data which involves scaling factors due to the signal processing. The plots show a typical comparison between the RAM locked an unlocked performance for the two different cases described.}
   \label{fig2}
\end{figure*}

RAM can be measured and actively reduced by determining the power falling on a fast photodiode at the modulation frequency and using that signal to correct the principal axes of the crystal using a feedback loop. The relative phase shift due to birefringence of the crystal is $\Delta\phi = k l (n_e-n_o)$, where $k \equiv 2\pi/\lambda_p$ is the wave number of the probe light with wavelength $\lambda_p$, $l$ is the length of the crystal, and $n_e$ ($n_o$) is the refractive index of the extraordinary (ordinary) wave. The phase shift can be externally controlled because $n_e$ and $n_o$ depend on the electric field present inside the crystal via the Pockels effect. Phase changes can then be due to a control electric field applied to the crystal, $\Delta\phi^{\mathrm{dc}}$, or from unwanted variations as mentioned in the introduction, $\Delta\phi^{\mathrm{n}}$. The correction phase needed to compensate the RAM can be determined by the condition that all odd harmonics of $\omega_m$ of the laser power falling on a photodiode placed after the EOM are zero for perfect phase modulation (alignment of the probe electric field vector along one of the principal axes of the crystal). For odd harmonics of $\omega_m$, the measured photocurrent will be \cite{Wong85}
\begin{equation}
I(n \omega_m) = -|E_0|^2 \sin(2\alpha)\sin(2\beta) J_n(M) \sin(n \omega_m t)\sin(\Delta\phi^{\mathrm{n}} + \Delta\phi^{\mathrm{dc}}), \label{photocurrent}
\end{equation}
where $n$ is an odd integer for $n>0$. $E_0$ is the E-field amplitude of the probe laser. $\alpha$ and $\beta$ are the relative angles of the GTP polarizer and analyzer with respect to the crystal axes, respectively. $J_n (M)$ is the $n^{th}$-order Bessel function with $M \equiv (\delta_e-\delta_o)$ defined as the difference between the modulation indices of the ordinary and extra-ordinary waves. $\delta_{o,e}$ depends on $n_{o,e}$ and the RF E-field applied to the EOM \cite{Wong85}. Pure phase modulation is realized by satisfying the condition $\sin(\Delta\phi^{\mathrm{n}} + \Delta\phi^{\mathrm{dc}})$ = 0. To correct for the RAM and obtain ideal phase modulation, a fast photodiode is used to measure the photocurrent given in Eqn.~\ref{photocurrent} by demodulating it at $\omega_m$ and actively feeding back this error signal into the EOM via the DC port of a bias tee as shown in Fig.\ref{fig1}. A PID loop based on a field programmable gate array is used to generate the signal that is applied to the bias tee \cite{schwett2011}. The laser power used for the RAM correction setup can be relatively large. Typically we used $\sim 200\,\mu$W.

A miniature Rubidium atomic clock (Microsemi SA.35m) is used to synchronize the three function generators used to control the modulation and demodulation in the setup. One function generator is used for the RF modulation of the EOM. The other two function generators are used as local oscillators (LO) for the RAM servo and the probe FM signal. The time drift of the clock in a 24 hour period is $<$ 7$\mu$s. The demodulation of the error signal for the RAM servo and the FM signal are performed using two double-balanced mixers. A band-pass filter (BPF) is used to suppress signal at other frequencies, such as higher order sidebands. The phase of the LO for the RAM servo is adjusted to obtain a maximum error signal from the mixer output. The phase of the LO used to demodulate the FM signal can produce in-phase or quadrature phase probe laser transmission signals \cite{Bjorklund80}.

\section{ Results and discussion}
%We observe that the change in RAM signal is linearly proportional to the applied DC voltage. %100 mV RAM signal is generated with 1v of DC voltage.
%We observe an improvement in the probe signal to noise $\geq$ 3 dB with the RAM Lock on. %Fig. 2 (a) shows the RAM signal vs time with RAM lock on (red curve) and off (black curve).

The stability of the RAM and the probe transmission signal can be studied in a log-log plot of the Allan deviation versus sampling time. The Allan deviation $\sigma_y(\tau)$ is the square root of the Allan variance defined as $\sigma^2_y(\tau) = \langle (y_{i+1}(\tau) -y_i(\tau))^2 \rangle/2 $, where $y_i(\tau)$ is the $i^{th}$ average fluctuation over the sampling time $\tau$. Fig.~\ref{fig2}(a) shows an example of the Allan deviation of the RAM signal plotted against sampling time with the RAM lock on (red curve) and off (black curve). The inset of Fig.~\ref{fig2}(a) shows the corresponding RAM signal as a function of time with the RAM lock on (red) and off (black). The modulation frequency was $10\,$MHz with a modulation depth $+8\,$dBm. The LO modulation depth was $+14\,$dBm. The Allan deviation is larger for short sampling times but decreases as the sampling time increases due to averaging. At larger sampling times, the Allan deviation of the unlocked signal increases due to slow drifts, such as polarization changes caused by temperature changes of the optics. The Allan deviation of the locked signal continues to decrease over the entire range of sampling times. The decrease in both traces shown in Fig.~\ref{fig2}(a) is characteristic of electronic noise, sometimes referred to as flicker phase noise, as well as noise due to the probe laser locking, white phase and frequency noise. Fig.~\ref{fig2}(a) shows that the RAM is largely eliminated and the FM sidebands are stable over long intervals of time due to the active stabilization of the FM sidebands.

 \begin{figure}[htbp]
  \begin{center}
 \scalebox{0.26}{\includegraphics{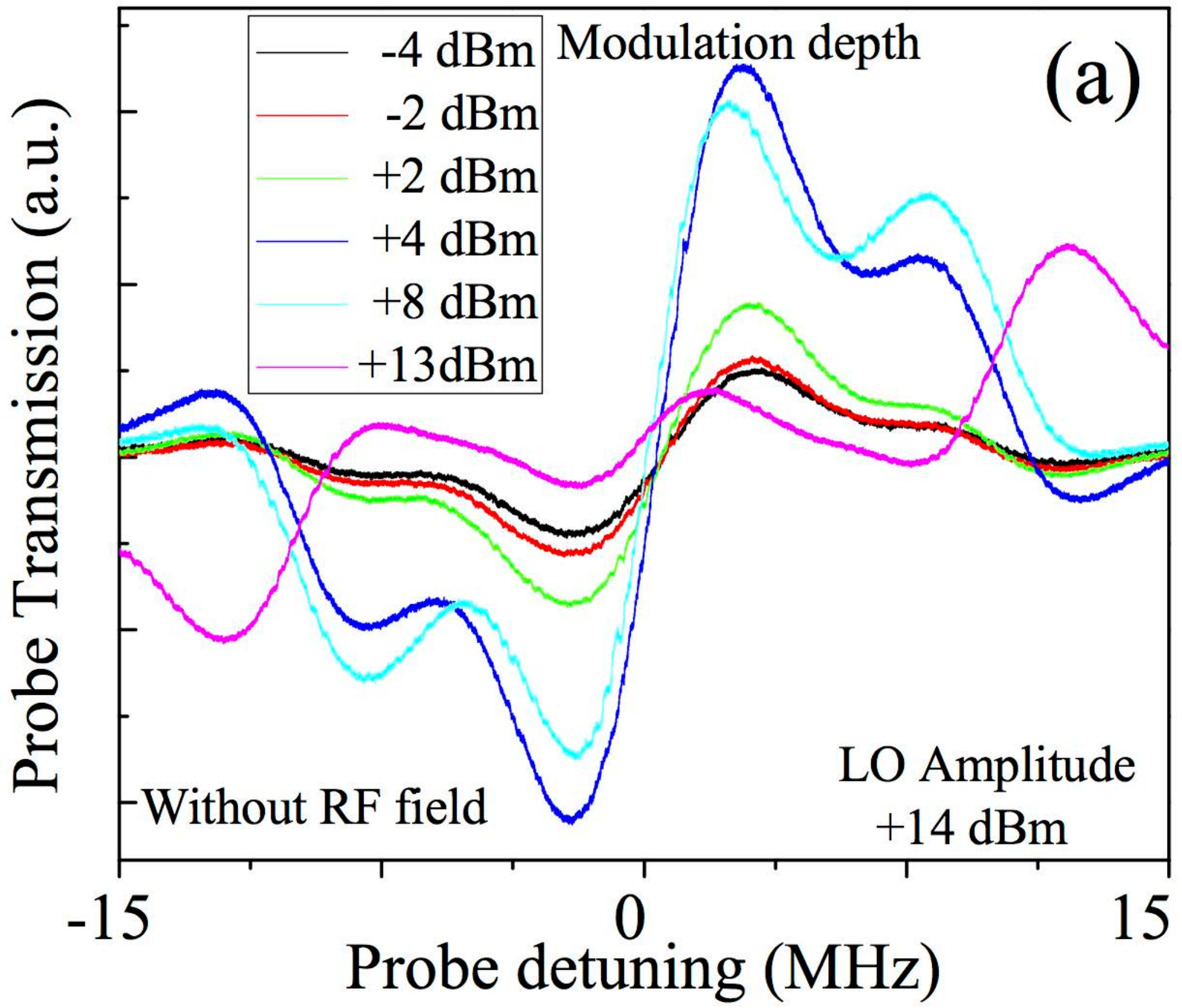}}
  \scalebox{0.45}{\includegraphics{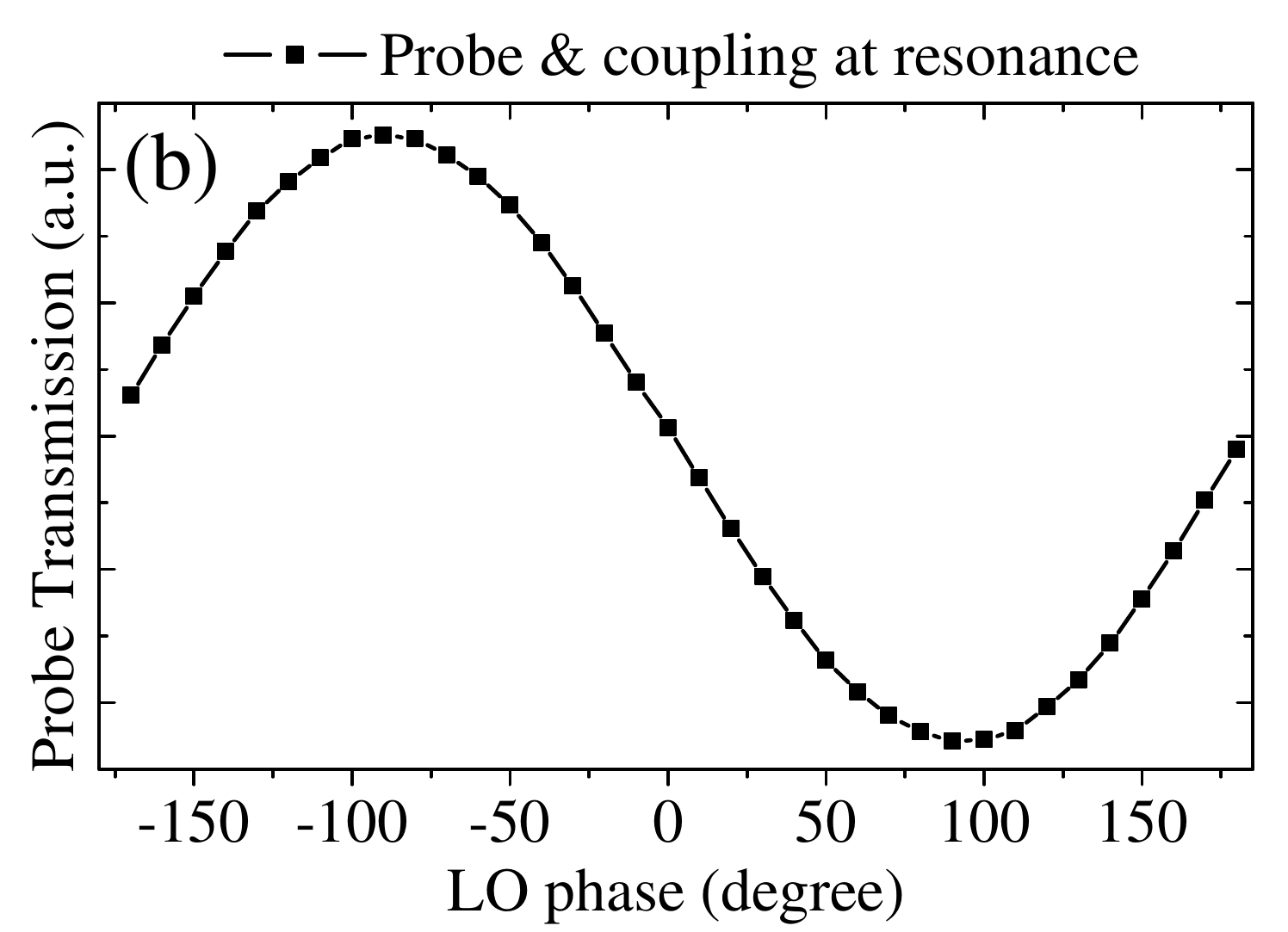}}   % requires the graphicx packageQCMC
  \end{center}
   \caption{(a) Shows the probe transmission as a function of probe laser detuning for different modulation depths. (b) Shows the probe transmission signal as a function of the LO phase with the probe and coupling lasers on resonance. The probe transmission is recorded under conditions where the reference phase used to demodulate the the probe transmission signal is in the quadrature condition.}
   \label{fig3}
\end{figure}

To further illustrate the point, Fig.~\ref{fig2}(b) shows an example of the Allan deviation of the probe transmission signal acquired using FM spectroscopy as a function of sampling time at two-photon resonance, the EIT resonance, with the RAM lock on (red curve) and off (black curve). The inset of Fig.~\ref{fig2}(b) shows the corresponding probe transmission signal in the time domain with RAM lock on (red curve) and off (black curve). For the EIT signal, we used $45\,\mu$W of probe laser power and $13\,$mW of coupling laser power. The probe laser beam had a diameter $1.5\pm 0.01\,$mm while the coupling beam diameter was $0.16\pm 0.01\,$mm. The corresponding probe and coupling laser Rabi frequencies are $2 \pi \times 5.6\pm 0.05\,$MHz and $2 \pi \times 5.7 \pm 0.05\,$MHz, respectively. Note that this is a fundamentally different measurement than shown in Fig.~\ref{fig2}(a) because now the signal depends on other parameters like the density of atoms in the vapor cell and the laser detunings. Consequently, we observe similar but not identical behavior as that shown in Fig.~\ref{fig2}(a). At first, $\tau < 10\,$s, both the locked and unlocked Allan variances increase. After the initial increase, the Allan deviation decreases with sampling time. For longer sampling times, the Allan deviation starts increasing again. The $\tau > 10\,$s behavior in this example is characteristic of white noise from electronics and then, most likely, slow environmental drifts of which there are more, since the experiment is more complicated than the measurement of the RAM on a single photodiode. For $\tau < 10\,$s, both curves most likely show random walk frequency noise due to the laser locking. Fig.~\ref{fig2}(b) demonstrates how sensitive these measurements are, particularly with respect to stabilizing the frequency and intensity of the lasers, and the utility of using the Allan deviation to determine the noise sources.

\begin{figure}[htbp]
\begin{center}
\includegraphics[width=2.2in]{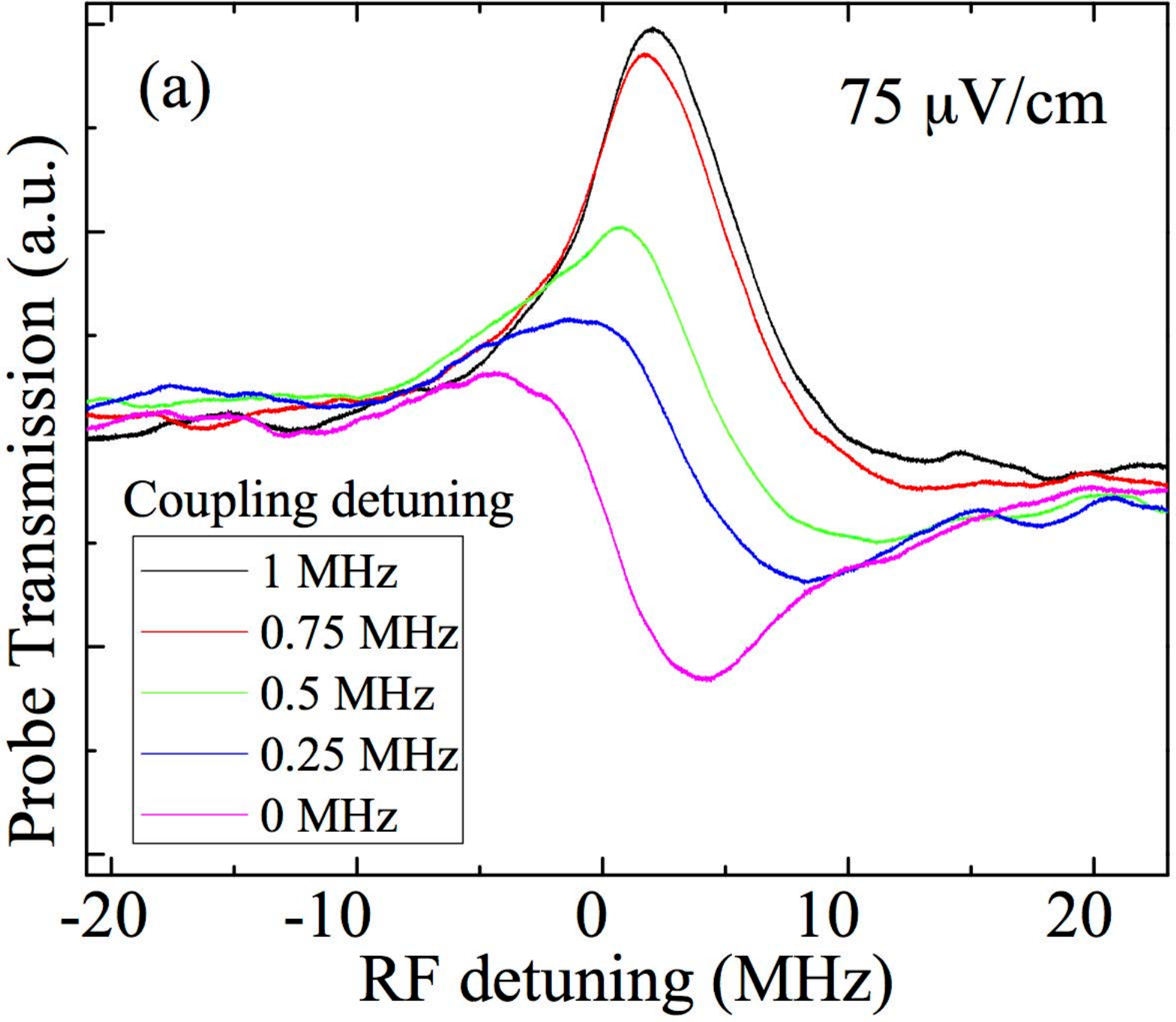} \includegraphics[width=2.2in]{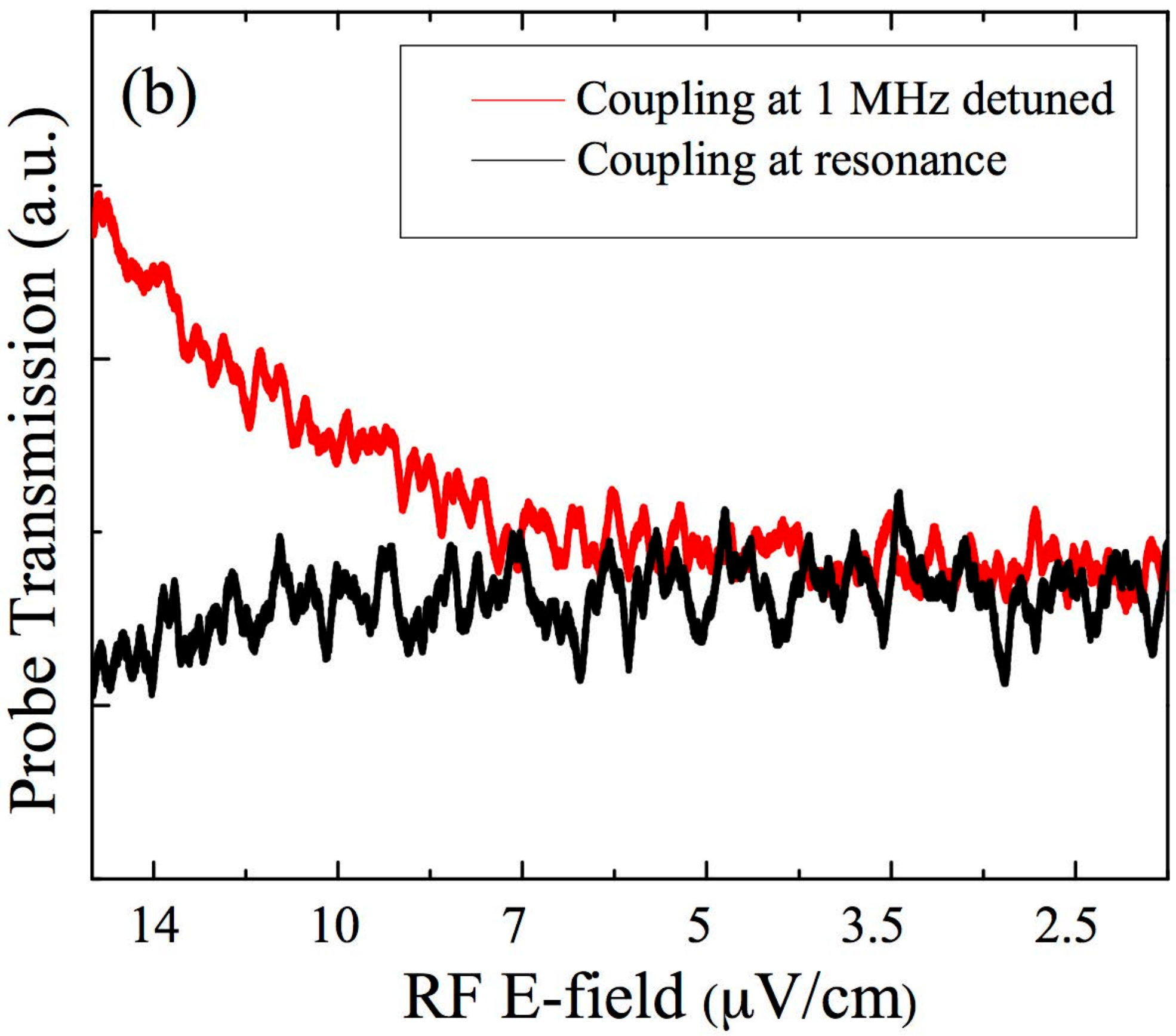} % requires the graphicx packageQCMC
  \end{center}
   \caption{(a) Shows the probe transmission as a function of RF detuning at an RF electric field strength of $75\,\mu$V/cm. The probe laser is resonant while the coupling laser is detuned. (b) Probe laser transmission as a function of RF electric field strength for the coupling laser on resonance (black curve) and detuned by 1 MHz (red curve). The data was taken with a $1\,$Hz detection bandwidth.}
   \label{fig4}
\end{figure}

\begin{figure}[htbp]
  \begin{center}
{\includegraphics[width=2.5in]{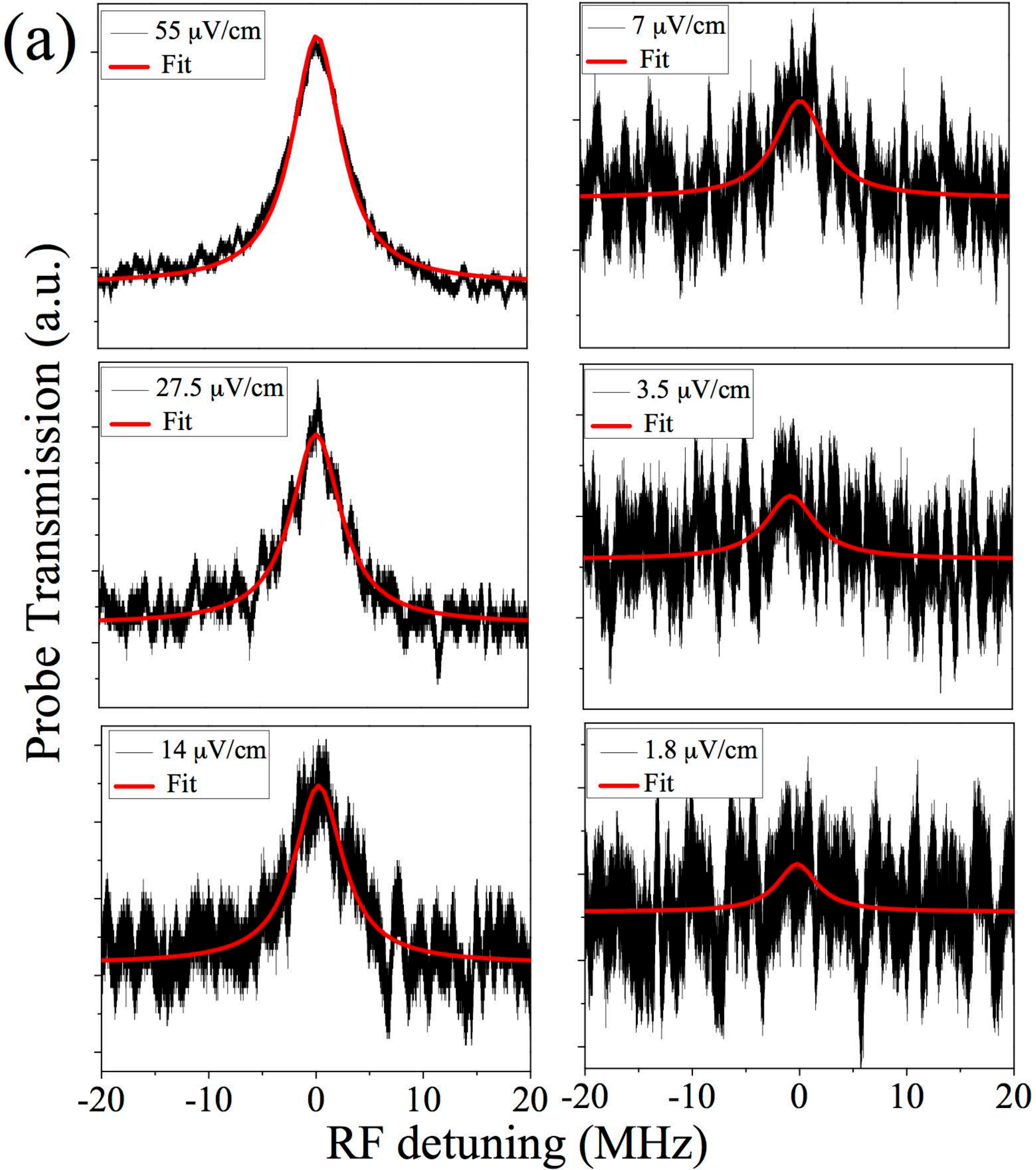}}  {\includegraphics[width=2.5in]{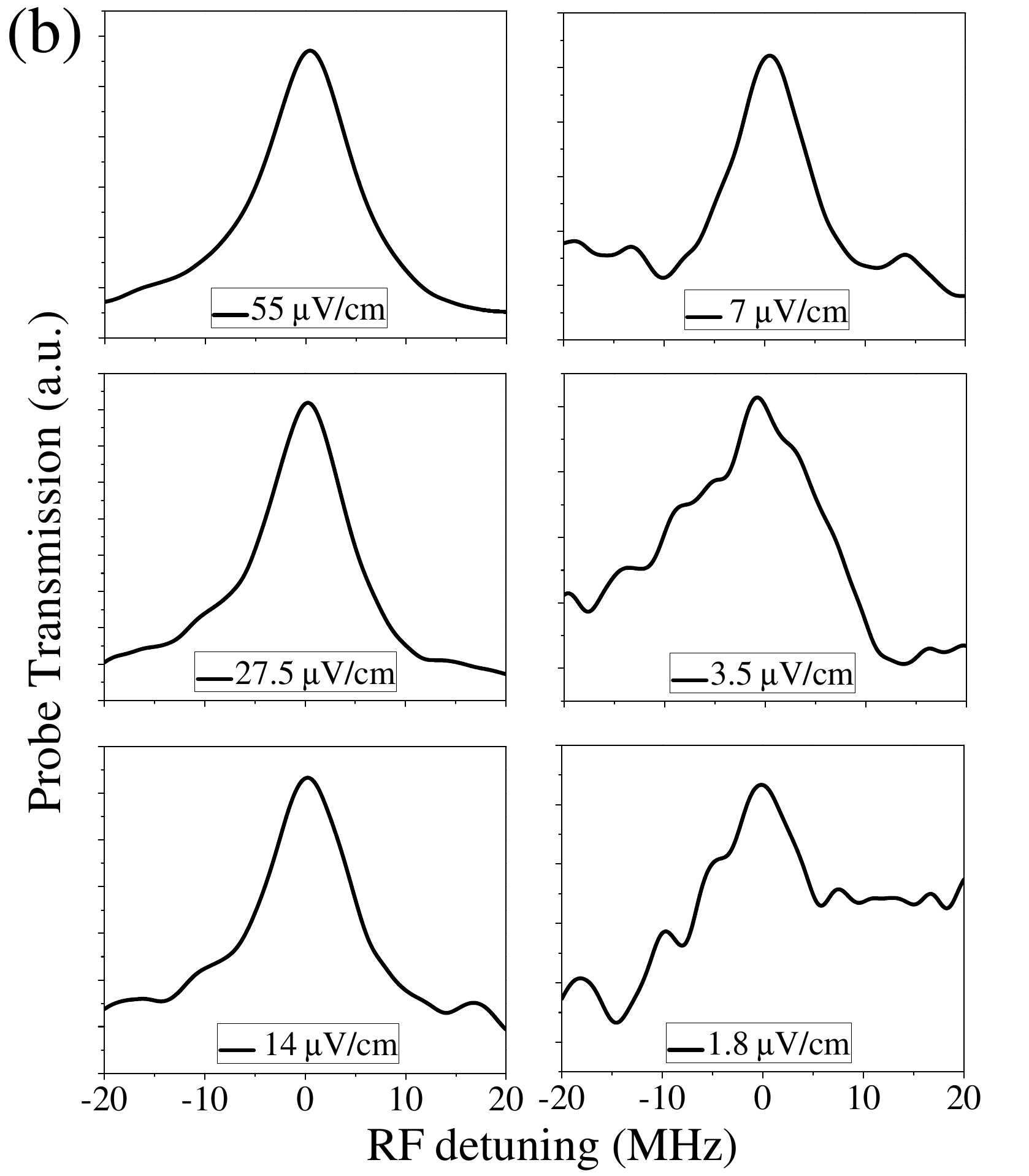} }% requires the graphicx packageQCMC\,~\,
  \end{center}
   \caption{(a) The plot shows probe transmission vs RF detuning at different values of RF E-field strengths. The black curves are the recorded data's and red curves are the Lorentzian fits. The data's were taken with same the probe and coupling Rabi frequencies $2 \pi \times 6.7 \pm 0.05\,$MHz and $2 \pi \times 7.0 \pm 0.05\,$MHz, respectively. (b) The plot shows probe signal vs RF detuning using matched filter at different values of RF E-field strengths. All parameters are same as (a). The data was acquired using a $1\,$Hz detection bandwidth. Each Lorentzian fit gave a full width at half maximum of $5.5\,$MHz.}
   \label{fig5}
\end{figure}

We studied the effect of the modulation depth on the probe laser transmission as a function of probe detuning in order to optimize the probe laser transmission signal. Fig.~\ref{fig3}, shows the probe transmission signal obtained with FM spectroscopy for the three-level ladder system shown in Fig.~\ref{fig0} with no RF E-field present. The slope of the probe transmission signal close to the probe resonance depends on the modulation frequencies and modulation depth. For a small modulation depth, the amplitude of the FM sideband is very small compared to the carrier wave. As we increase the modulation depth, the amplitude of the sideband grows larger which increases the slope of the FM signal. For a large modulation depth, we also observe that the amplitude of the higher-order harmonics grows larger and the slope of the FM signal decreases \cite{Supplee94}. To optimize the slope of the FM probe laser transmission signal at a modulation frequency of $10\,$MHz, we used a modulation depth of $+8\,$dBm for the EOM and a LO modulation depth of $+14\,$dBm to demodulate the FM signal. Fig.~\ref{fig3}(b) shows the probe transmission as a function of the LO phase at the EIT probe transmission resonance. For quadrature LO phases, we observe a antisymmetric FM spectral lineshape about the line center. The steepness of the probe transmission is maximum at these quadrature phases consistent with prior work \cite{Bjorklund80}. We use the quadrature phase of the FM derived probe laser transmission to perform the RF E-field measurements shown in Fig.~\ref{fig4} and Fig.~\ref{fig5}.

We also optimized the detuning of the coupling laser so that the applied RF E-field would cause the maximum change in probe laser transmission. Fig.~\ref{fig4}(a) shows probe laser transmission as a function of RF detuning for several different coupling laser detunings. The probe laser is tuned to resonance. We used $65\,\mu$W of probe laser power and $20\,$mW of coupling power which corresponds to Rabi frequencies of $2 \pi \times 6.7 \pm 0.05\,$MHz and $2 \pi \times 7.0 \pm 0.05\,$MHz, respectively. In Fig.~\ref{fig4}(a), the RF E-field strength is $75\,\mu$V$\,$cm$^{-1}$. The figure shows that the amplitude of the probe transmission is optimized for a detuned coupling field for our EIT parameters and modulation frequency. At $1\,$MHz coupling field detuning, we observed maximum probe transmission.  Fig.~\ref{fig4}(b) shows the probe laser transmission plotted against RF E-field strength for two different coupling detunings. Fig~\ref{fig4}(b) illustrates the effect of optimizing the coupling detuning for the measurements. One can observe from the graph that the probe transmission as a function of RF E-field amplitude is more sensitive for the $1\,$MHz coupling laser detuning. %The plot shows that the change in the probe transmission is large for the case of a 1 MHz detuned coupling field which corresponds to the peak of the steep signal for the probe scan.

%at the peak of the dispersive FM signal with respect to probe detuing

Fig.~\ref{fig5}(a) shows weak RF E-field measurements performed by detecting the probe laser transmission as a function of RF E-field detuning at a $1\,$MHz coupling laser detuning. The probe laser is resonant. The Rabi frequencies of the probe and coupling lasers are the same as in Fig.~\ref{fig4}. The RF E-field amplitude was calibrated by measuring it in the AT regime using the Rydberg atom-based sensor and extrapolating the power reading on the RF generator used to drive the antenna \cite{Sedlacek12}. In Fig.~\ref{fig5}(a), the black curve shows the measured probe laser transmission while the red curve shows a Lorentzian fit to the lineshape. We used a Lorentzian fit because the lineshape is expected to be dominated by power broadening since we increased the probe laser power to optimize the signal with regards to photon shot noise and lineshape. The full width at half maximum was $5.5\,$MHz for each fit within the fitting error. The Lorentzian fit the data curves much better than a Gaussian particularly in the wings of each peak. Fig.~\ref{fig5}(b) shows the same plots as Fig.~\ref{fig5}(a) after processing the data using a matched filter.  Matched filtering is useful for improving the SNR when the form of the signal is known. It detects or extracts a known signal with high sensitivity that has been contaminated by noise \cite{Riris94,McDonough95,Vasilyev15}. To implement the matched filter, we convolve a Lorentzian distribution function,
\begin{equation}
F(\nu,\sigma,A,\nu_c) =\frac{A \sigma}{(\nu-\nu_c)^2 + \sigma^2},
\end{equation}
with the measured signal, where $\sigma$ is the full width at half maximum, $A$ is the amplitude, and $\nu_c$ is the line center. The matched filter improves the visibility of the signal, but the signal to noise level is approximately the same at the smallest E-field amplitudes, presumably because photon shot noise is limiting the sensitivity. The results in Fig.~\ref{fig5}a show a sensitivity $\sim 3\,\mu$V$\,$cm$^{-1}\,$Hz$^{-1}$. Each data point corresponds to a $\sim 1\,$Hz detection bandwidth. With matched filtering, one can detect a RF electric field amplitude of $\sim$ 1.8 $\mu$V$\,$cm$^{-1}$.

The current sensitivity achieved with FM spectroscopy is about three orders of magnitude worse than what can be achieved for an optimized projection noise limit of the atomic sensor \cite{Fan15}. For the parameters used in this effort, the projection noise limit is about 20 times better than what we demonstrated. The minimum RF E-field that an atomic sensor can measure depends on the RF transition dipole moment, $\mu_{RF}$; the effective number of atoms participating, $N$; and the dephasing time, $T_{2}$ \cite{Fan15}. The atomic projection noise limited sensitivity of the Rydberg atom-based E-field sensor for $N$ atoms can be calculated as%The shot noise limited RF E-field sensitivity of the atoms can be calculated as
\begin{equation}
\frac{E^{min}_{RF}}{\sqrt{Hz}} = \frac{h}{\mu_{RF} \sqrt{N T_2}},
\label{eq3}
\end{equation}
where $h$ is the Planck's constant. For the parameters used in our experiment; $\mu_{RF}= 1745~e a_{0}$, $N \approx 10^5$ participating atoms and $T_2 = 0.5\,\mu$s, Eqn.~\ref{eq3} gives a projection noise limited sensitivity of $\sim 160\,$nV$\,$cm$^{-1}\,$Hz$^{-1/2}$. For $N$, we made a conservative estimate which takes into account the Doppler averaging and relative Rydberg excitation of the atomic sample based on a density matrix calculation of the system shown in Fig.~\ref{fig0}a. The result of the projection noise limited sensitivity estimate shows that the shot noise of the atoms in the vapor cell is not limiting the sensitivity, since it is a factor of 20 better than what was measured in the experiments. As in our prior work, photon shot noise on the photo-detector is limiting the sensitivity \cite{Kumar16}. The photon shot noise limited SNR of a detector is $\sqrt{2 \eta e^2 P_\mathrm{s} \Delta f/ h\nu}$, where $\eta$ is the quantum efficiency, $\Delta f$ is the detection bandwidth, $P_\mathrm{s}$ is the power falling on the detector, and $\nu$ is the frequency of the light \cite{Boyd}. For the experiments, a total probe laser power of $\sim 65\,\mu$W fell on the detector, but less than $10\%$ carried the signal because of the size of the laser beam overlap, necessitated in our setup by the small dipole moment of the coupling laser transition, and modulation.   For a $1\,$Hz detection bandwidth, as used in the prior analysis, the shot noise limited SNR is $\sim 2 \times 10^6$ using this reasoning. The smallest probe transmission signals shown in this paper are $\sim 0.1\%$ of the EIT signal while the overall EIT signal without the RF electric field is $\sim 0.1\%$ of the absorption signal. These estimates show that photon shot noise is the primary noise source for low probe laser transmission signals. Surpassing the photon shot noise limit in cases where sub-$100\,$nV$\,$cm$^{-1}$Hz$^{-1/2}$ sensitivities need to be achieved is nontrivial for Rydberg atom-based RF E-field sensing because increasing the probe laser power leads to power broadening as well as the possible reduction of $T_2$ depending on how the other parameters of the sensor are constrained \cite{Fan15}.

%The suppression of residual Doppler effect using three photon EIT readout has the potential to improve the sensitivity and accuracy of the method by widening the AT regime because a peak splitting is easier to measure than a change in peak height. We have predicted that the three photon readout for the Cs $6S_{1/2}\leftrightarrow 6P_{1/2} \leftrightarrow 9S_{1/2} \leftrightarrow 53P_{3/2}$ system  with RF coupling $53P_{3/2}\leftrightarrow 52D_{5/2}$ can achieve a sensitivity of $\sim 500\,$nV$\,$cm$^{-1}\,$Hz$^{-1/2}$ for a measurement of the peak splitting and an amplitude change detection sensitivity of $\sim 200\,$nV$\,$cm$^{-1}\,$Hz$^{-1/2}$ \cite{Kumar16}. Details of the three photon calculations will be the subject of a forthcoming article. The RAM canceled FMS with matched filtering will play a vital role to achieve such detection sensitivity for RF E-field measurements in three photon readout method. One approach to reduce the photon shot noise is to use squeezed light for the probe transition \cite{Li16,Andersen16}. The atomic shot noise limited performance would allow us to make absolute E-field measurements, for example, of cosmic microwave background radiation \cite{Trangsrud12} and tool for radio astronomy \cite{Wilson09}. To reduce the noise floor, we can carefully choose the combination of detector and amplifier.

\section{Conclusion}
We have shown that FM spectroscopy with active control of RAM improves the readout SNR of Rydberg atom-based RF E-field sensing when compared to our prior approach \cite{Sedlacek12}. We have achieved a sensitivity of $\sim$ 3 $\mu$Vcm$^{-1}$Hz$^{-1/2}$ and demonstrated the detection of weak RF E-fields, $\sim$ 1.8 $\mu$V$\,$cm$^{-1}$, using matched filtering. This sensitivity limit is the same as what we achieved recently utilizing a MZI \cite{Kumar16}. The current experimental sensitivity of the RF E-field measurement is worse than the atomic projection noise limit for realistic cases where $N$ and $T_2$ are not too small because of photon shot noise. It is possible to change the parameters of the system to incrementally improve the sensitivity, however, photon shot noise is a barrier that must be overcome to achieve optimal projection noise limited sensitivity. The MZI read-out method was similarly bounded because the laser Rabi frequencies are determined by a desire to maximize coherence times and avoid power broadening. The agreement between the FM spectroscopy and MZI results supports the idea that photon shot noise is an important source of noise to address if the sensitivity of Rydberg atom-based RF E-field measurement is to be significantly improved. Some approaches that we are investigating to surpass the photon shot noise limit are to utilize non-resonant two-photon excitation or squeezed light for the probe transition. FM spectroscopy is very versatile and can be used in a portable, compact package that is less complicated than an interferometer but can reach similar performance levels. RF E-field measurements at sensitivities below the photon shot noise limit can possibly be used in a wide range of applications in medical science, such as detection of early stage breast cancer\cite{Rosen02}, and as a tool for radio astronomy to detect absolute levels of thermal background radiation and explore fundamental physics of the universe \cite{Wilson09, Bernardis13}. At present, Rydberg atom-based sensing is well-poised to become the standard for electric field measurement, particularly useful for calibration of RF and terahertz devices, from GHz to THz.

%\section*{Funding}To achieve the sensitivity below the photon shot noise limit, one approach is to use squeezed light for the probe laser.

\section*{Acknowledgments}

The authors would like to thank Jiteng Sheng for the fruitful discussions. This work was supported by the DARPA Quasar program by a grant through the ARO (60181-PH-DRP). We also acknowledge support from the National Reconnaissance Office. HK acknowledges support from the Carl-Zeiss Foundation.

\end{document}